\newcommand{\Pl}{\partial}
\newcommand{\ts}{\textstyle}
\newcommand{\fder}[2]{\frac{{\ts d \/ #1}}{{\ts d\/ #2}}}
\newcommand{\fpar}[2]{\frac{{\ts \Pl \/ #1}}{{\ts \Pl \/ #2}}}
\newcommand{\nder}[3]{\frac{{\ts d^{#1} \/ #2}}{{\ts d \/ #3^{#1}}}}

\newcommand{\bee}{\begin{equation}}
\newcommand{\ene}{\end{equation}}
\newcommand{\beea}{\begin{eqnarray}}
\newcommand{\enea}{\end{eqnarray}}
\newcommand{\sech}{\mbox{sech$\,$}}
\documentclass[preprint,aps,showpacs]{revtex4}
\usepackage{dcolumn}
\usepackage{graphicx}
\usepackage{latexsym}
\usepackage{amsmath}
\usepackage{amssymb}
\begin{document}
\title{Jeans Instability in a viscoelastic fluid}
\author{ M. S. Janaki, N. Chakrabarti and D. Banerjee}{
\affiliation{ Saha Institute of Nuclear Physics, I/AF Bidhannagar,
Calcutta 700 064, India}

\begin{abstract}
The well known Jeans instability is studied for a viscoelastic, gravitational fluid using generalized hydrodynamic equations of motions.
It is found that the threshold for the onset of  instability appears at  higher wavelengths in a viscoelastic medium.
Elastic effects playing a role similar to thermal pressure are found to lower the growth rate of the gravitational instability.
Such features may manifest themselves in matter consituting dense astrophysical objects.
\end{abstract}
\maketitle
\section{Introduction}
In astrophysical scenarios, the simplest theory that describes the aggregation
of masses in space is the Jeans instability.  The system comprises of particles
that can aggregate together depending on the relative magnitude of the gravitational
force to pressure force.  Whenever the internal pressure of a gas is too weak to balance the self-gravitational
force of a mass density perturbation, a collapse occurs. Such a mechanism was first
studied by Jeans\cite{kn:jeans}. The Jeans' instability is of central importance in
understanding the process of formation of stars, planets, comets, asteroids and other astrophysical objects.

 Several works investigating the properties of Jeans instability in dusty plasmas\cite{kn:prs} have appeared in literature, considering the presence of massive charged dust grains
in astrophysical fluids\cite{kn:bingham}, and have lead to interesting results due to contributions from both gravitational and electrostatic  forces.
In the context of dense astrophysical objects such as the interior of super dense white dwarfs and the atmospheres of neutron stars,
extensive studies of  Jeans instability  have been carried out  by taking into account the  role of quantum\cite{kn:shukla}
as well as  non-ideal effects\cite{kn:ren}.  The role of magnetic field in arresting the Jeans collapse has also been studied\cite{kn:verheest}-\cite{kn:cramer}.
A central idea in the study of the  instability by including various factors is to find ways of arresting the gravitational collapse.
All the studies  are mostly based on hydrodynamic models in presence of viscous, buoyant
as well as gravitational forces, and are applicable in the context of flowing viscous  matter that is the constituent of all main sequence stars.
On the other hand, superdense degenerate star matter\cite{kn:bastrukov} can be regarded to be made up of solid matter
possessing properties of a viscoelastic medium with the macroscopic motions governed by laws of solid mechanics.
It is well known from theories of continuum mechanics that the elastic behaviour of
solids is manifested by  shear oscillations.
The observation of torsional oscillations
in white dwarfs points to the possible existence  of elastic behaviour in such
environment.
It is our supposition that in the transition stage between the viscous liquid state
and the elastic solid state, the characteristics of stellar matter are similar to that of
a viscoelastic fluid where both the properties work together.  An appropriate model to
study such a viscoelastic regime is the generalized hydrodynamic model\cite{kn:frenkel}-\cite{kn:kaw}.  The model  treats
the normal fluid viscosity coefficient as a viscoelastic operator.
In the present work, we would like to investigate the effects  of  viscoelasticity on a self-gravitating fluid,
in particular to the Jeans instability.

In Section -II we present the generalized hydrodynamic model containing the contribution of the
gravitational and pressure gradient force terms used to describe a medium that is capable
of supporting viscoelastic stresses.  This equation is supplemented by the equations of
continuity, Poisson's equation for gravitational potential and equation of pressure
for adiabatic behaviour of fluid flows.  For an infinite, homogeneous fluid
in the strongly coupled fluid limit, elastic stresses contribute to fluid thermal pressure to arrest gravitational condensation.  In section III, we describe the analysis of
gravitational instability by treating a fluid  with variable density.  For a kind of
self-consistent equilibrium profile that we choose, it is found that the instability
disappears for higher mode numbers.
\section{Equilibrium equations and stability analysis for a fluid with uniform density}
 In the present study we consider a neutral fluid which is strongly coupled so that viscosity and elasticity
act on the same footing. We will use the generalized hydrodynamic model  to treat the visco-elastic
property.  In a
    viscoelastic medium the normal viscosity coefficient behaves as a viscoelastic operator
    as described in Frenkel's book \cite{kn:frenkel}.
    We follow the same procedure and write the generalized
    equation of motion for  a viscoelastic medium
    as
    \begin{equation}
    \left(1+ \tau \fder{}{t}\right)\left[\rho \fder{\bf v}{t}
    +\rho \nabla \psi + c_{s}^{2} \nabla \rho \right ] = \eta \nabla^{2} {\bf v}
    + (\zeta+\frac{\eta}{3})\nabla (\nabla \cdot {\bf v})
    \label{ghe}
    \end{equation}
    where $\rho$ is the mass density, ${\bf v}$ is the fluid velocity, $\psi$ is the gravitational potential and  $c_s$ is the sound speed.
    The viscoelastic properties of the medium are characterized by  the relaxation time $\tau$
     \cite{kn:frenkel},
     shear viscosity $\eta$   and
     bulk viscosity coefficient $\zeta$.  The evolution equation for mass density is described by the
     continuity equation.
 \bee
 \fpar{\rho}{t} + \nabla \cdot \left( \rho {\bf v} \right) = 0.
 \label{cnt}
 \ene
The Gravitational potential $\psi$ is related to the mass density through the Poisson's equation
\bee
\bigtriangledown^{2} \psi = 4 \pi G \rho .
\label{pn}
\ene
In a Newtonian fluid, the role of viscosity terms is to give rise to the damping of sound modes.  However, a viscoelastic medium governed by Eqs.(\ref{ghe}) and (\ref{cnt}) supports the
propagation of both longitudinal and transverse viscoelastic modes\cite{kn:db} in the limit  $\omega \tau  >> 1$.
This limit physically implies that the mode frequency is much larger than the inverse of the visco-elastic relaxation time.

   For a homogeneous neutral fluid, there is an ambiguity in defining the equilibrium.
   The concept of Jeans swindle \cite{kn:chandrasekhar}  has been used  in the local dispersion relation, to avoid the zero order gravitational field.
   The homogeneous plasma is described by the constant variables
     $\rho=\rho_0$, ${\bf v} = 0,$.
     With the equilibrium  mentioned above  we perturbed the system with a small
     amplitude disturbance i.e. ${\bf v}= 0+ {\bf v}_1 ({\bf r},t)$,
     $ \psi = \psi_0 +  \psi_1({\bf r},t)$ and  $ \rho =\rho_0 + \rho_1 ({\bf r},t)$
     where all the variables with subscript one are perturbations.
     Linearizing  Eqs. (\ref{ghe}), (\ref{cnt})  and (\ref{pn})
     around the equilibrium mentioned above we have

 \bee
    \left(1+ \tau \fpar{}{t}\right)\left[\rho_0 \fpar{{\bf v}_1}{t}
    + \rho_{0} \nabla \psi_{1} + c_{s}^{2} \nabla \rho_1  \right ] = \eta \bigtriangledown^{2} {\bf
    v}_1
    + (\zeta+\frac{\eta}{3})\nabla (\nabla \cdot {\bf v}_1)
    \label{ghe1}
    \ene

\bee
 \fpar{\rho_{1}}{t} + \nabla \cdot \left( \rho_{0} {\bf v_{1}} \right) = 0
 \label{cnt1}
 \ene

\bee
\bigtriangledown^{2} \psi_{1} = 4 \pi G \rho_{1}
\label{pn1}
\ene
 Since the above equations are linear and the medium is homogeneous,  we can Fourier transform these equations
    assuming the solutions for the perturbed variables  in the form
    $\sim \exp[-i(\omega t-{\bf k}\cdot {\bf r})]$. Here $\omega$ is the frequency
and ${\bf k}$ is the wave vector of the mode under consideration.
Substituting the
perturbed solutions in Eqs. (\ref{ghe1}) - (\ref{pn1}) we find
\bee
    \left(1-i\omega \tau \right)\left[-i \omega \rho_0 {\bf v}_1
    + i\rho_{0} {\bf k} \psi_{1} + i c_{s}^{2} {\bf k}  \rho_{1}  \right ] = - \eta k^{2}
{\bf v}_1 - (\zeta+\frac{\eta}{3}) {\bf k} ({\bf k} \cdot {\bf v}_1)
    \label{ghe2}
    \ene
\bee
- \omega\rho_{1} +  \rho_{0} \left( {\bf k}\cdot{\bf v_{1}} \right) = 0
 \label{cnt2}
 \ene
\bee
 k^{2} \psi_{1} = - 4 \pi G \rho_{1}
\label{pn2}
\ene
Taking dot product of {\bf k} on both sides of Eq. (\ref{ghe2}) and using the Eqns. (\ref{cnt2})
 and (\ref{pn2}) we obtain
\bee
    \left(1-i\omega \tau \right)\left[\omega^{2} \rho_1
    + 4\pi G \rho_{0}  \rho_{1} - c_{s}^{2}  k^{2} \rho_{1}  \right ] = - i\omega \frac{\left(\zeta + \frac{4}{3}\eta \right) }{\rho_{0}} k^{2} \rho_{1}
    \label{ghe3}
    \ene
Now, in the limit $\omega \tau \gg 1$, from Eqn (\ref{ghe3}) we get the
dispersion relation
\bee
\omega^{2} = - \omega_{J}^{2} + c_{s}^{2} k^{2} + \frac{\left(\zeta + \frac{4}{3}\eta \right) }
{\tau \rho_{0}} k^{2}
\label{gin}
\ene
where $\omega_{J}^2 = 4 \pi G \rho_{0}$ is the Jean's frequency of the fluid. Eq. (\ref{gin}) constitutes the
linear dispersion relation describing Jean's instability for a homogeneous plasma.  The equation suggests that
the presence of viscoelastic effects in the medium  contribute to its stability against fluctuations in gravitational potential.  In the next section, we consider a more realstic case of a non uniform equilibrium with a
zero order gravitational field.\\

\section{Stability analysis with non uniform mass density}
Before going to the stability analysis it is useful to explain the
     equilibrium solutions.
    For this we consider the case of a fluid with a non uniform density distribution where for simplicity
 we model the density variation to be present in one direction only.
First, we write  equilibrium equations related to  density and gravitational potential  which can be used
in the stability analysis.
 In equilibrium, viscosities are assumed to be small then ,  Eqs. (\ref{ghe}) and (\ref{pn}) describing a system with non uniform mass density and gravitational potential are given by

\bee
\rho_0 \fder{\psi}{z} = - c^{2}_{s} \fder{\rho_{0}}{z}
\label{ne1}
\ene
and
\bee
\nder{2}{\psi}{ z} = 4 \pi G \rho_{0}
\label{ne2}
\ene
Combining these two equations we get a differential equation for the normalized equilibrium density

\bee
\frac{d^{2} \ln \hat{\rho_{0}}}{dz^{2}} + \frac{2}{\lambda_{J}^{2}} \hat{\rho_{0}} = 0,
\label{ne3}
\ene
where $\lambda_{J}^{2} = {c_{s}^{2}}/{2\pi G \rho_{00}}$ and $\rho_{00}$ is constant density.
The above equation (\ref{ne3}) enables us to find the following equilibrium density and
gravitational field distribution as
\bee
\rho_{0} = \rho_{00} \sech^{2}\dfrac{z}{\lambda_{J}},\;\;
\psi= - c_s^2 \ln \rho_0,\;\;\; g=-\frac{c_s^2}{\lambda_J} \tanh\left(\frac{z}{\lambda_J}\right),
\ene

For the linear stability analysis we will perturb the system around these inhomogeneous
solutions.
The  continuity, momentum and Poisson's equations after linearization become
\bee
 \fpar{\rho_{1}}{t} +  \left( {\bf v_{1}}\cdot \nabla \right)\rho_{0}+ \rho_{0}
 \left( \nabla \cdot {\bf v_{1}}\right) = 0
 \label{lcnt1}
\ene
\bee
\left( 1 - i \omega \tau \right) \left[ -i \omega \left( \nabla \cdot {\bf v_{1}} \right) + \nabla^2\psi_1
 + {c^{2}_{s}}\bigtriangledown^{2}\frac{\rho_{1}}{\rho_{0}}\right]
= \frac{\left( \zeta + \frac{4}{3} \eta \right)}{\rho_{0}} \bigtriangledown^{2}\left( \nabla \cdot {\bf v_{1}}\right)
\label{lghe1}
\ene
\bee
\nabla^2\psi_1  =  4 \pi G \rho_{1}
\label{lpn1}
\ene
In the above equations equilibrium variables are inhomogeneous in $z$, therefore  are all the perturbed quantities are assumed
 to be of the form
$ f({\bf r},t) \sim  f(z) \exp(i(- \omega t + {\bf k}_{{\bot}}\cdot {\bf r}))$
where f({\bf r},t) are taken to be $\rho_{1}({\bf r}, t), {\bf v}_{1}({\bf r}, t)$ and
${\bf k}_\perp={\hat e}_x k_x +{\hat e}_y k_y$, $k_x, k_y$ are wave vectors along $x,y$ directions.

 After a straightforward algebra Eqs. (\ref{lcnt1}) and (\ref{lpn1})
 reduces to
\bee
\left(1+\frac{M^{2}}{\sech^{2}\hat{z}}\right) \nder{2}{n}{\hat z}-2 \tanh \hat{z} \fder{n}{\hat z}+
\left[ 2 \frac{\omega^{2}}{\omega_{J}^{2}}+2 \sech^{2}\hat{z}- k^{2}_{\perp} \lambda_{J}^{2}\left(1+\frac{M^{2}}{\sech^{2}\hat{z}} \right) \right]n=0
\label{nl1}
\ene
where
\[
 n= \frac{\rho_1}{\rho_0(z)}, \;\;\;v_{c}^{2} = \frac{\zeta + \frac{4}{3} \eta}{\rho_{0} \tau},\;\;\;
   \rho_{0}(z)= \rho_{00} \sech^{2}\hat{z},\;\;\; M=\frac{v_c}{c_s}.
\]
In the above equation, $v_c$ and $M$ denote the velocity of compressional viscoelastic mode and Mach number respectively.
To derive above equation we have assumed that in $z$ direction perturbed pressure gradient is much larger that the
perturbed gravitational potential gradient.

 First, in the simplest level we attempt  solutions of eq.(\ref{nl1}) for $M=0$ i.e. in absence of viscoelastic effect. In this case an exact analytical solutions are available. Substituting $M=0$, eq.(\ref{nl1}) reduces to
\bee
 \nder{2}{n}{\hat z}-2 \tanh \hat{z} \fder{n}{\hat z} + \left[2 \frac{\omega^{2}}{\omega_{J}^{2}}+
2 \sech^{2}\hat{z}-k^{2}_{\perp} \lambda_{J}^{2}\right]n=0
\label{nl0}
\ene
The above Eq. (\ref{nl0}) has an exact solution of the form\cite{kn:fricke}:
\bee
\rho_1 = \rho_{00} {(\sech^2 \hat z)}^{(1+\sqrt{17})/4},
\label{solex}
\ene
and  the corresponding dispersion relation given by
\bee
2\frac{\omega^{2}}{\omega^2_J} =   k_\perp^{2}\lambda_J^2 + \frac{(\sqrt{17}-7)}{2}=k_\perp^{2}\lambda_J^2 -1.4
\label{dp0}
\ene
If we compare this nonlocal dispersion relation with the local dispersion relation given in Eq.(\ref{gin}), the threshold wavenumber for istability is reduced by a factor of 0.6.

Exact analytical solution of
Eq. (\ref{nl1}) is not possible. Therefore we solve Eq. (\ref{nl1}) numerically  by representing it in the form of a eigenvalue problem with $\omega$ as the eigenvalue.  The positive imaginary part
$\gamma$ of the eigenvalue (if at all exists) gives the growth rate of the instability.  For the range of $k_\perp \lambda_J$ values studied, only one pair of imaginary eigenvalues were obtained that correspond to growing and damped modes.
In fig.1, we have shown the  eigen solutions $\rho_1$ (unnormalized) of  Eq. (\ref{nl0}) corresponding to the growing mode plotted against  $z/\lambda_J$.
The exact analytical solution given in Eq.(\ref{solex}) is also shown in the same figure. The exact match between the two solutions is a verification of the numerical scheme
adapted by us for solving Eq. ({\ref{nl1}).
\begin{figure}[h]
             \includegraphics[width=3.0in,height=3.0in]{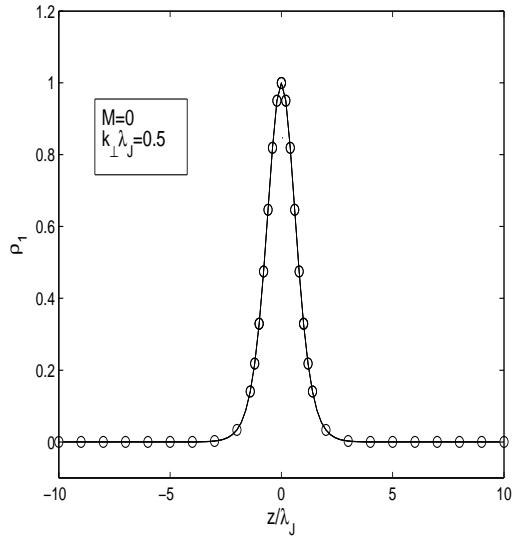}
             \caption{Plot of $\rho_1$ against $z/\lambda_J$ for $M=0$  and $k_\perp\lambda_J=0.5$
              with the continuous line showing the numerical solution of Eq. (\ref{nl0}) and the `o' line indicating the corresponding analytical solution.}
             \label{x1}
             \end{figure}
Fig. 2 shows the  growth rate $\gamma/\omega_J $  plotted against $k_\perp \lambda_J $
as obtained from Eq. (\ref{dp0}) as well as from the numerical solution of Eq. (\ref{nl0}).

\begin{figure}[h]
             \includegraphics[width=3.0in,height=3.0in]{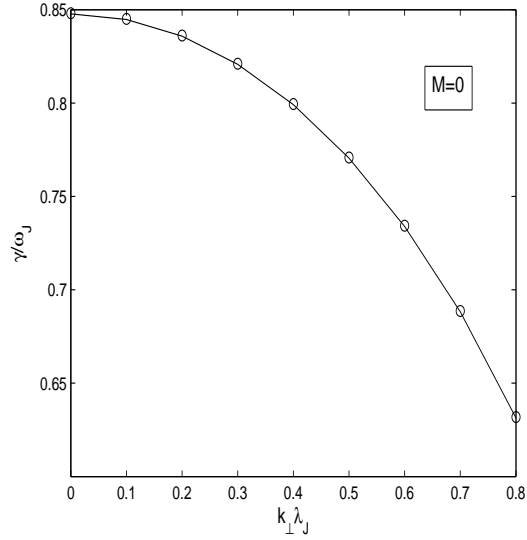}
             \caption{Normalized Jeans instability growth rate for $M=0$  as a function of
              $k_\perp\lambda_J$. The continuous line indicates the numerical solution while `o' line indicates the growth rate obtained from exact analytical result. }
             \label{x2}
             \end{figure}

  It is already mentioned that for non zero values of $M$,  an exact analytical solution of Eq. (\ref{nl1})is not possible.
It is found that the normalized   solution of Eq. (\ref{nl1}), $\hat {\rho_1}$ has the nature of a $\sech {\hat z} $ profile and
 can be fitted with an analytical profile of the form $(\sech\hat {z})^\alpha$ with $\alpha$ depending on the value of $M$.
In Fig 3 (a), we have shown the $\hat \rho_1$ profile for $M=0.4$ with $k_\perp \lambda_J =0.5$ for the growing mode. together with an analytical fit
with $\mu = 0.52$. In fig. 3(b) we have plotted the corresponding unnormalized solution $\rho_1$ against $z/\lambda_J$.
\begin{figure}[hl]
            \includegraphics[width=5.0in,height=1.8in]{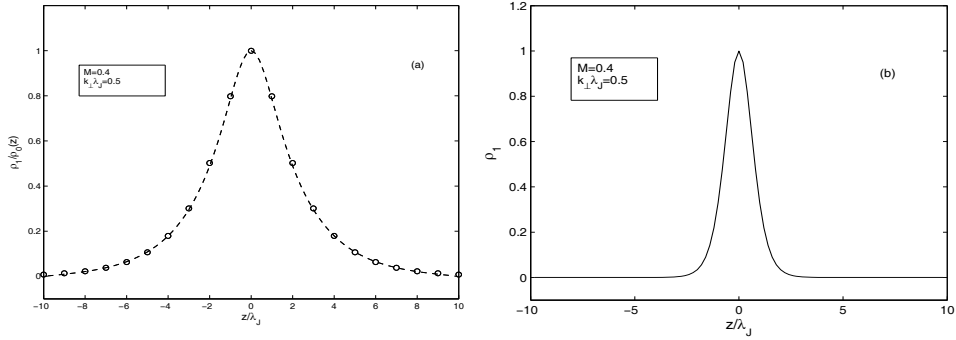}
             \caption{(a) Plot of $n$ against $z/\lambda_J$ for $M=0.4$  and $k_\perp\lambda_J=0.5$ with
              `---' line showing the normalized solution $\hat \rho_1$ and the `o' line  representing a
 $(\sech\hat {z})^{0.52}$  profile  (b) Plot of $\rho_1$ with the same parameters as used in (a).}
             \label{x3}
             \end{figure}
\begin{figure}[ht]
             \includegraphics[width=3.0in,height=3.0in]{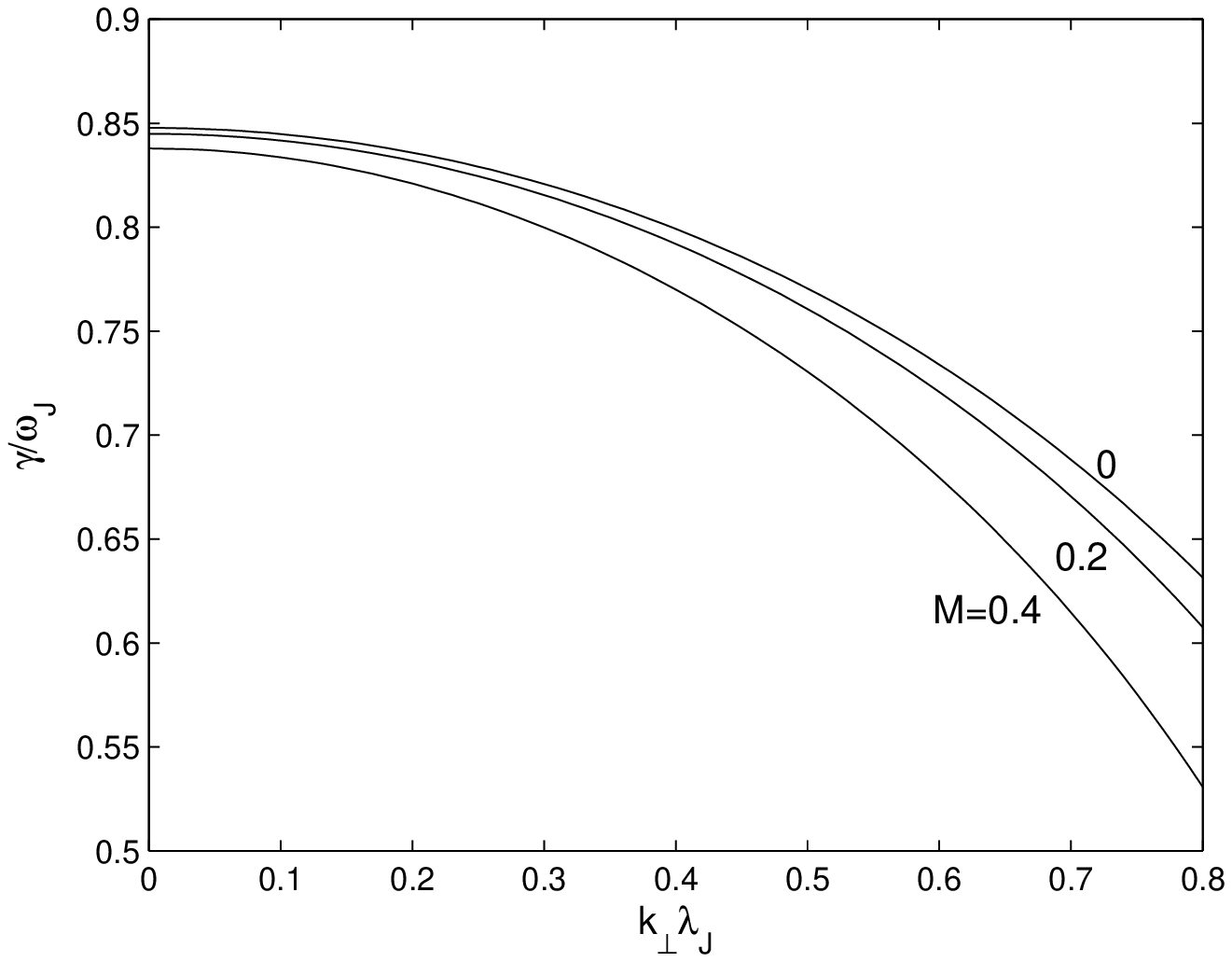}
             \caption{Normalized Jeans instability growth rate for different values of $M$  as a function of
              $k_\perp\lambda_J$.}
             \label{x4}
             \end{figure}
The growth rates for different values of $M$ are shown plotted against $k_\perp\lambda_J$, showing
that with increase in values of $M$, the growth rate decreases.

We also attempt to obtain an approximate solution for Eq. (\ref{nl1}).
Let us consider $\xi = \tanh\hat{z}$ and take $\xi \ll 1$ so that higher powers of $\xi$ can
be neglected,

\bee
\nder{2}{n}{\xi} - 2 \xi \left( \frac{M^{2}+2}{M^{2}+1} \right) \fder{n}{\xi}+ \left[\frac{2+2 \overline{\omega}^{2}- {k}_{\perp}^{2}\lambda_J^2(1+M^{2})]}{(1+M^{2})}\right]n = 0
\label{nl2}
\ene
To solve this equation we can  use  the transformations
\bee
n = N(\xi) e^{\int \frac{M^{2}+2}{M^{2}+1}\xi d\xi}~~{\rm{and}}~~ \bar \xi = \xi \sqrt{\frac{M^{2}+2}{M^{2}+1}}
\ene
then Eq. (\ref{nl2}) reduced to the well known  Hermite's equation
\bee
 \nder{2}{N}{\bar \xi} + \left[\frac{M^{2} + 4 + 2 ({\omega}^{2} /{\omega_J^2}) - k^{2}_{\perp}\lambda_J^2(1+M^{2})}{M^{2}+2}  - \bar \xi^{2}\right] N = 0.
\label{nl4}
\ene
The  solutions of the Hermite equations are found as
$N(\bar \xi) = \exp{(-\frac{\bar \xi^{2}}{2})} H_{\nu}(\bar \xi)$
where $H_{\nu}$ is a Hermite polynomial.  A physically acceptable solution can only be found
if it satisfies the dispersion relation
\bee
\frac{M^{2} + 4 + 2 ({\omega}^{2} /{\omega_J}^{2})- k^{2}_{\perp}\lambda_J^2(1+M^{2})}{M^{2}+2} = 2\nu + 1
\label{fnal}
\ene
Different solutions can be obtained for different values of $\nu$.
For $\nu = 0$, we get back homogeneous density dispersion relation $\omega^{2} = - \omega^{2}_{J} + k_\perp^{2}c_s^2(1+ M^{2})$.
For $\nu=1$, we obtain the dispersion relation
$\omega^{2} = \left( 1+ M^{2}\right) \left( k_\perp^{2}c^{2}_{s} + \omega^{2}_{J}\right) $ that is satisfied for real values of $\omega$ for any value of the wavenumber.
Similar results are obtained for all higher values of $\nu$ showing that there is only one mode corresponding to $\nu=0$ that gives rise to imaginary eigenvalues and the instability disappears for higher mode numbers. This feature is also reflected from the numerical analysis of Eq.(\ref{nl1}) where only one mode was found to be unstable.
\section{Conclusions }
A study of Jean's instability has been carried out for a viscoelastic fluid that exhibits effects of both viscosity and elasticity  using generalized hydrodynamic equations of motion.
For a Newtonian fluid it is well known that self-gravity leads to an instability for all wavenumbers $k< 4\pi G\rho_0/c_s^2 $.  For a viscoelastic
fluid, the upper limit of wavenumber, upto which the gravitational instability is observed  is lowered.  For a given value of perpendicular wavenumber, the growth rate is shown to decrease
with increase in the values of elastic modulus coefficients. Such results have been obtained both for the idealized uniform density case as well as for a one-dimensional equilibrium density
profile of the $\sech^2$ form.  These results may be relevant for stellar matter that is known to exhibit viscoelastic behaviour.
%

\end{document}